\documentclass[conference]{IEEEtran}
\usepackage{mathrsfs}
\usepackage{cite}
\usepackage{amsmath}
\usepackage{amssymb}
\usepackage{stfloats}
\usepackage{graphicx}
\usepackage{setspace}
\usepackage{xcolor}

\ifCLASSOPTIONcompsoc
\usepackage[tight,normalsize,sf,SF]{subfigure}
\else
\usepackage[tight,footnotesize]{subfigure}
\fi

\begin{document}
\title{Device-to-device Cooperation in Massive MIMO Systems with Cascaded Precoding}
%
\author{
\authorblockN{Yinsheng~Liu$^{\dag}$, Geoffrey Ye Li$^{\ddag}$, and Wei Han$^{*}$}
\authorblockA{$^{\dag}$State Key Laboratory of Rail Traffic Control and Safety, Beijing Jiaotong University, Beijing, China, e-mail: ys.liu@bjtu.edu.cn.\\
$^{\ddag}$ School of ECE, Georgia Institute of Technology, Atlanta, USA, e-mail: liye@ece.gatech.edu.\\
$^{*}$ Huawei technologies, Co. Ltd., China, e-mail: wayne.hanwei@huawei.com.\\}
}

\maketitle

\begin{abstract}
This paper investigates user cooperation in massive multiple-input multiple-output (MIMO) systems with cascaded precoding. The high-dimensional physical channel in massive MIMO systems can be converted into a low-dimensional effective channel through the inner precoder to reduce the overhead of channel estimation and feedback. The inner precoder depends on the spatial covariance matrix of the channels, and thus the same precoder can be used for different users as long as they have the same spatial covariance matrix. Spatial covariance matrix is determined by the surrounding environment of user terminals. Therefore, the users that are close to each other will share the same spatial covariance matrix. In this situation, it is possible to achieve user cooperation by sharing receiver information through some dedicated link, such as device-to-device communications. To reduce the amount of information that needs to be shared, we propose a decoding codebook based scheme, which can achieve user cooperation without the need of channel state information. Moreover, we also investigate the amount of bandwidth required to achieve efficient user cooperation. Simulation results show that user cooperation can improve the capacity compared to the non-cooperation scheme.
\end{abstract}


\section{Introduction}
As a promising technique for the next generation cellular systems, massive multiple-input multiple-output (MIMO) has gained a lot attention recently \cite{FRusek,LLu}. By installing a huge number of antennas at the base station (BS), massive MIMO can significantly increase the spectrum- and energy-efficiencies of wireless networks.\par

In MIMO systems, the downlink channel state information (CSI) at the BS can help improve the performance significantly. In regular MIMO systems where the antenna number at the BS is relatively small, downlink CSI can be first estimated at the user terminal and then fed back to the BS through limited feedback \cite{YLiu_survey,DJLove}. In massive MIMO systems, however, traditional channel estimation and feedback can be hardly used due to the large overhead caused by the huge number of antennas. To reduce the overhead, cascaded precoding has been proposed in \cite{AAdhikary,JNam} where the precoder is divided into an inner precoder and an outer precoder. The inner precoder converts the high-dimensional physical channel into a low-dimensional effective channel such that traditional channel estimation and feedback can be used directly with respect to the effective channel.\par

The inner precoder depends on the spatial covariance matrix of the channel \cite{AAdhikary}. The same precoder can be thus used for different users as long as they have the same spatial covariance matrix. Spatial covariance matrix is determined by the surrounding environment of the user terminal. Measurement results in \cite{IViering} have shown that the spatial covariance matrix is very stable over time, which can be translated into stability over space since the time variation of wireless channel is essentially caused by the motion of the terminal over space \cite{RHClarke}. In other words, if the users are close to each other, they will share the same surrounding environment and thus the same spatial covariance matrix. In this situation, it is possible to achieve user cooperation by sharing receiver information through some dedicated link, such as device-to-device (D2D) communication \cite{AAsadi,DFeng}. Note that D2D communications in our scenario are different from the traditional one because (a) only two users are considered as a transmission pair in traditional D2D while more-than-two users are allowed in our scenario, and (b) traditional D2D communications cannot happen if there is no need of data transmission between the users while the D2D communications in this scenario are to share receiver information for user cooperation and it thus can happen even if no need of data transmission. Therefore, the D2D communications in our scenario will be called as \emph{D2D cooperation} to distinguish from the traditional one.\par

\begin{figure*}
  \centering
  \includegraphics[width=4.5in]{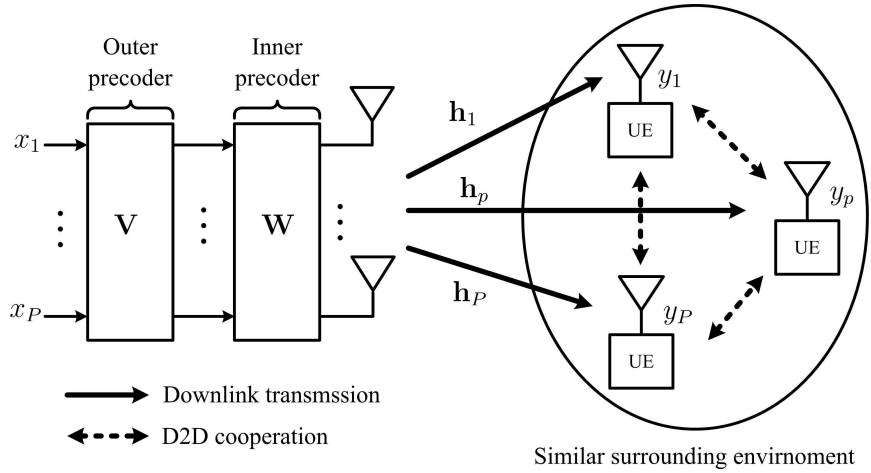}\\
  \caption{System model for cascaded precoding in massive MIMO systems with D2D cooperation.}\label{system}
\end{figure*}

Similar to traditional D2D, D2D cooperation can exploit either the licensed or unlicensed spectrums \cite{AAsadi}. In both cases, the amount of receiver information that needs shared should be as small as possible to save spectrum resources. For this purpose, we will develop a decoding codebook based scheme, which can achieve user cooperation without the need of CSI. In particular, a decoding codebook is first pre-stored by the BS and the users, respectively. The BS picks up a decoding matrix from the decoding codebook and informs users the selection result. Each user will then use the selected decoding matrix for demodulation based on the shared receive signal information (RSI). In practice, the RSI should be first quantized and then sent to the other users via specific D2D cooperation link, which will not only result in an extra quantization noise but also consume extra cooperation bandwidth. Based on the connection between the quantization noise and the cooperation bandwidth, we will find out how much cooperation bandwidth is required to achieve satisfied performance.\par

The rest of this paper is organized as follows. The system model is introduced in Section II. The user cooperation scheme and the corresponding analysis are presented in Section III. In Section IV, we will discuss the consumption of the cooperation bandwidth. Finally, simulation results and conclusions are in Section V and Section VI, respectively.

\section{System Model}

As in Fig.~\ref{system}, we consider a BS with $M$ transmit antennas and $P$ users each with a single receive antenna. $\mathbf{W}\in\mathbb{C}^{M\times D}$ denotes the inner precoder where $D$ ($D\geq P$) is the dimension of the effective channel, and $\mathbf{V}\in\mathbb{C}^{D\times P}$ denotes the outer precoder. From Fig.~\ref{system}, the received signal vector, $\mathbf{y}=(y_1,\cdots,y_P)^{\mathrm{T}}$, can be represented by
\begin{align}\label{recv_signal}
\mathbf{y}=\mathbf{H}^{\mathrm{H}}\mathbf{W}\mathbf{V}\mathbf{x}+\mathbf{z},
\end{align}
where $\mathbf{x}=(x_1,\cdots,x_P)^{\mathrm{T}}$ and $\mathbf{z}=(z_1,\cdots,z_P)^{\mathrm{T}}$ with $x_p$ and $z_p$ indicating the transmitted signal from the $p$-th user and the corresponding additive noise with $\mathrm{E}(|x_p|^2)=1$ and $\mathrm{E}(|z_p|^2)=N_0$, and $\mathbf{H}=(\mathbf{h}_1,\cdots,\mathbf{h}_P)$ denotes the channel matrix with $\mathbf{h}_p$ indicating the $M\times 1$ channel vector of the $p$-th user.\par

Measurement results in \cite{IViering} have shown that the spatial covariance matrix is very stable over time, which can be also translated to stability over space since the time variation of wireless channel is essentially caused by the motion of the terminal over space \cite{RHClarke}. In other words, the spatial covariance matrix can remain constant within a spatial area (with size about $30$ meters according to \cite{IViering}). In this sense, we can assume that the spatial covariance matrices are the same for different users if they are close to each other, that is
\begin{align}
\mathbf{R}=\mathrm{E}(\mathbf{h}_p\mathbf{h}_p^{\mathrm{H}})~\mathrm{for}~p=1,2\cdots,P.
\end{align}
From \cite{AAdhikary}, the inner precoder is composed of the eigen-vectors of $\mathbf{R}$ corresponding to the $D$ largest eigenvalues. In this situation, the received signal in (\ref{recv_signal}) can be rewritten as
\begin{align}\label{recv_signal_eff}
\mathbf{y}=\mathbf{H}_{\mathrm{e}}^{\mathrm{H}}\mathbf{V}\mathbf{x}+\mathbf{z},
\end{align}
where $\mathbf{H}_{\mathrm{e}}=\mathbf{W}^{\mathrm{H}}\mathbf{H}\in\mathbb{C}^{D\times P}$ denotes the low-dimensional effective channel.\par

The signal model with respect to the effective channel in (\ref{recv_signal_eff}) is actually similar to the traditional downlink multiuser MIMO \cite{DTse}. In traditional multiuser MIMO systems, users can distribute over the entire cell and thus each user is not aware of the other users' positions. As a result, traditional user cooperation approaches have to develop extra transmission scheme or protocols to achieve information sharing among the users \cite{ASendonaris,HJu,TEHunter}. In our scenario, different users will share the same spatial covariance matrix and thus they should be very close to each other. In this situation, information sharing can be achieved through D2D communication based cooperation, which has widely accepted as a key technology in future wireless systems \cite{AAsadi}.

\section{User Cooperation}
In this section, we will first describe the deocding codebook based approach for user cooperation, and then present the corresponding analysis will be presented.
\subsection{Decoding Codebook}
To save the amount of information that needs to be shared, only the RSI, $y_p$'s, are shared by the cooperated users. In this section, we assume the RSI can be perfectly shared so that each user can have a copy of the received signal vector, $\mathbf{y}$. In this situation, the transmitted signal to the $q$-th user can be recovered at the $q$-th user terminal as
\begin{align}\label{coop_recv}
\widehat{x}_p&=\mathbf{q}_p^{\mathrm{H}}\mathbf{y}\nonumber\\
&=\mathbf{q}_p^{\mathrm{H}}\mathbf{H}_{\mathrm{e}}^{\mathrm{H}}\mathbf{V}\mathbf{x}+\mathbf{q}_p^{\mathrm{H}}\mathbf{z},
\end{align}
where $\mathbf{q}_p$ denotes the decoding vector for the $q$-th user. Rewrite (\ref{coop_recv}) in a matrix form with $q=1,2,\cdots,P$, we have
\begin{align}
\widehat{\mathbf{x}}=\mathbf{Q}^{\mathrm{H}}\mathbf{H}_{\mathrm{e}}^{\mathrm{H}}\mathbf{V}\mathbf{x}+\mathbf{Q}^{\mathrm{H}}\mathbf{z},
\end{align}
where $\mathbf{Q}=(\mathbf{q}_1,\mathbf{q}_2,\cdots,\mathbf{q}_P)$ indicates the decoding matrix, which is a $P\times P$ unitary matrix selected at the BS from a pre-determined decoding codebook,
\begin{align}
\mathbb{Q}=\{\mathbf{Q}_1,\mathbf{Q}_2,\cdots,\mathbf{Q}_{2^b}\},
\end{align}
where $b$ denotes the number of quantization bits for the decoding codebook. For simplicity, we consider a random codebook in this paper as in \cite{NRavindran}, where each codeword is constructed by generating a $P\times P$ matrix with complex Gaussian entries and then forming the codeword through eigenvalue decomposition.\par

To select a specific codebook from the codebook, BS needs to know the effective CSI, $\mathbf{H}_{\mathrm{e}}$, which is assumed to be perfectly known by accurate channel estimation at the user terminals and feedback. Then, the overall channel observed will be $\mathbf{H}_{\mathrm{e}}\mathbf{Q}$. If non-cooperative zero-forcing (ZF) precoder is used as the outer precoder with respect to the overall channel, the signal-to-noise ratio (SNR) for the $p$-th user will be
\begin{align}\label{SNR}
\mathrm{SNR}_p=\frac{1}{N_0\{[\mathbf{Q}^{\mathrm{H}}\mathbf{H}_{\mathrm{e}}^{\mathrm{H}}\mathbf{H}_{\mathrm{e}}\mathbf{Q}]^{-1}\}_{(p,p)}},
\end{align}
where $\mathbf{A}_{(p,p)}$ denotes the $(p,p)$-th entry of matrix $\mathbf{A}$. The SNR expression in (\ref{SNR}) is actually similar to that in \cite{DJLove_Unitary} where the limited feedback based precoder design has been investigated. In this sense, the selection of a decoding matrix can be viewed as the duality of selecting a precoding matrix. It is shown in \cite{DJLove_Unitary} that the upper bound of the SNR is determined by the minimum eigenvalue of the overall channel matrix and thus the codeword should be selected to maximize the minimum eigenvalue. However, the upper bound based approach in \cite{DJLove_Unitary} cannot be used in our scenario because $\mathbf{Q}$ is a unitary matrix and thus $\mathbf{H}_{\mathrm{e}}\mathbf{Q}\mathbf{Q}^{\mathrm{H}}\mathbf{H}_{\mathrm{e}}^{\mathrm{H}}=\mathbf{H}_{\mathrm{e}}\mathbf{H}_{\mathrm{e}}^{\mathrm{H}}$. It means any codeword in the codebook will result in the same eigenvalue.\par

Alternatively, we consider a selection criterion based on the original SNR rather than the upper bound of the SNR. In particular, we select the codeword, $\mathbf{Q}_{\mathrm{o}}$, to maximize the average SNR, that is
\begin{align}\label{ave_SNR}
\mathbf{Q}_\mathrm{o}&=\operatorname*{arg~max}_{\mathbf{Q}\in\mathbb{Q}}\frac{1}{P}\sum_{p=1}^P\mathrm{SNR}_p\nonumber\\
&=\operatorname*{arg~max}_{\mathbf{Q}\in\mathbb{Q}}\frac{1}{N_0P}\sum_{p=1}^P\frac{1}{\mathbf{q}_p^{\mathrm{H}}(\mathbf{H}_{\mathrm{e}}^{\mathrm{H}}\mathbf{H}_{\mathrm{e}})^{-1}\mathbf{q}_p},
\end{align}
where we have used the identity $\mathbf{Q}^{\mathrm{H}}=\mathbf{Q}^{-1}$ for the second equation since $\mathbf{Q}$ is a unitary matrix. Once the optimal decoding matrix is determined, each user can demodulate its corresponding data symbols through (\ref{coop_recv}).

\subsection{Analysis}


In this subsection, we will provide an insightful analysis on the effect of the quantization bits.\par
Denote
\begin{align}
\mathbf{H}_{\mathrm{e}}^{\mathrm{H}}\mathbf{H}_{\mathrm{e}}=\mathbf{U}\mathbf{\Lambda}\mathbf{U}^{\mathrm{H}},
\end{align}
where $\mathbf{U}=(\mathbf{u}_1,\cdots,\mathbf{u}_P)$ is the eigen-matrix of $\mathbf{H}_{\mathrm{e}}^{\mathrm{H}}\mathbf{H}_{\mathrm{e}}$ and $\mathbf{\Lambda}=\mathrm{diag}\{\lambda_1,\cdots,\lambda_P\}$ is the corresponding eigenvalue matrix. The denominator in (\ref{SNR}) can be then rewritten as
\begin{align}\label{denominator}
\mathbf{q}_p^{\mathrm{H}}(\mathbf{H}_{\mathrm{e}}^{\mathrm{H}}\mathbf{H}_{\mathrm{e}})^{-1}\mathbf{q}_p&=\sum_{i=1}^{P}\lambda_i^{-1}|\mathbf{q}_p^{\mathrm{H}}\mathbf{u}_i|^2\nonumber\\
&=\lambda_p^{-1}\cos^2\theta_{p,p}+\sum_{i\neq p}\lambda_i^{-1}\cos^2\theta_{p,i},
\end{align}
where $\theta_{p,i}=\arccos(\mathbf{q}_p^{\mathrm{H}}\mathbf{u}_i)$ denotes the angle between $\mathbf{q}_p$ and $\mathbf{u}_i$. Note that $\mathbf{q}$ is with unit norm and $\mathbf{u}_i$'s are orthogonal vectors, we therefore have
\begin{align}
\sum_{a=1}^P\cos^2\theta_{p,a}=\|\mathbf{q}\|_2^2=1.
\end{align}
As a result, $\cos^2\theta_{p,p}+\cos^2\theta_{p,i}\leq 1$ for any $1\leq i\leq P$, and therefore
\begin{align}\label{inequ2}
\cos^2\theta_{p,i}&\leq 1 - \cos^2\theta_{p,p}\nonumber\\
&= \sin^2\theta_{p,p}.
\end{align}
Using (\ref{inequ2}), an upper bound of (\ref{denominator}) can be obtained by
\begin{align}
&\mathbf{q}_p^{\mathrm{H}}(\mathbf{H}_{\mathrm{e}}^{\mathrm{H}}\mathbf{H}_{\mathrm{e}})^{-1}\mathbf{q}_p
\leq\lambda_p^{-1}\cos^2\theta_{p,p}+\sum_{i\neq p}\lambda_i^{-1}\sin^2\theta_{p,p}\nonumber\\
&=\lambda_p^{-1}+\{\mathrm{Tr}[(\mathbf{H}_{\mathrm{e}}^{\mathrm{H}}\mathbf{H}_{\mathrm{e}})^{-1}]-2\lambda_p^{-1}\}\sin^2\theta_{p,p},
\end{align}
where we have used the identity
\begin{align}
\mathrm{Tr}[(\mathbf{H}_{\mathrm{e}}^{\mathrm{H}}\mathbf{H}_{\mathrm{e}})^{-1}]=\sum_{p=1}^P\lambda_p^{-1}.
\end{align}


As a result, the mean of the lower bound for the average SNR can be obtained by
\begin{align}\label{inequ3}
&\mathrm{E}\{\overline{\mathrm{SNR}}\}=\frac{1}{N_0P}\sum_{p=1}^P\mathrm{E}\left\{\frac{1}{\mathbf{q}_p^{\mathrm{H}}(\mathbf{H}_{\mathrm{e}}^{\mathrm{H}}\mathbf{H}_{\mathrm{e}})^{-1}\mathbf{q}_p}\right\}\nonumber\\
&\geq\frac{1}{N_0P}\sum_{p=1}^{P}\mathrm{E}\left\{\frac{1}{\lambda_p^{-1}+\{\mathrm{Tr}[(\mathbf{H}_{\mathrm{e}}^{\mathrm{H}}\mathbf{H}_{\mathrm{e}})^{-1}]-2\lambda_p^{-1}\}\sin^2\theta_{p,p}}\right\},
\end{align}
where the expectation is with respect to $\sin^2\theta_{p,p}$. If denote
\begin{align}
f(\sin^2\theta_{p,p})=\frac{1}{\lambda_p^{-1}+\{\mathrm{Tr}[(\mathbf{H}_{\mathrm{e}}^{\mathrm{H}}\mathbf{H}_{\mathrm{e}})^{-1}]-2\lambda_p^{-1}\}\sin^2\theta_{p,p}},\nonumber
\end{align}
it can be verified that $f(\sin^2\theta_{p,p})$ is always a convex function with respect to $\sin^2\theta_{p,p}$ in the region determined by $\sin^2\theta_{p,p}>0,f(\sin^2\theta_{p,p})>0$. Therefore, using Jensen's inequality to (\ref{inequ3}) \cite{SBoyd}, we can obtain
\begin{align}\label{inequ4}
&\mathrm{E}\{\overline{\mathrm{SNR}}\}\geq\nonumber\\
&\frac{1}{N_0P}\sum_{p=0}^{P-1}\frac{1}{\lambda_p^{-1}+\{\mathrm{Tr}[(\mathbf{H}_{\mathrm{e}}^{\mathrm{H}}\mathbf{H}_{\mathrm{e}})^{-1}]-2\lambda_p^{-1}\}\mathrm{E}\{\sin^2\theta_{p,p}\}},
\end{align}

In above, the decoding vector, $\mathbf{q}_p$, is generated through a random codebook and it thus can be viewed as isotropically distributed over $\mathbb{C}^{P\times P}$. This is also the case for the eigenvector $\mathbf{u}_i$. Therefore, similar to the analysis in \cite{NJindal}, we have
\begin{align}\label{Jindalform}
\mathrm{E}\{\sin^2\theta_{p,p}\}\approx 2^{-\frac{b}{P-1}}.
\end{align}
The lower bound of SNR in (\ref{inequ4}) can be thus obtained as
\begin{align}\label{inequ5}
&\mathrm{E}\{\overline{\mathrm{SNR}}\}\geq\nonumber\\
&\frac{1}{N_0P}\sum_{p=0}^{P-1}\frac{1}{\lambda_p^{-1}+\{\mathrm{Tr}[(\mathbf{H}_{\mathrm{e}}^{\mathrm{H}}\mathbf{H}_{\mathrm{e}})^{-1}]-2\lambda_p^{-1}\}2^{-\frac{b}{P-1}}}.
\end{align}
From (\ref{inequ5}), the lower bound of the average SNR can be improved exponentially as the rising of the bit number. When $B\rightarrow \infty$, the lower bound of the average SNR will be
\begin{align}\label{inequ5}
\frac{1}{P}\sum_{p=1}^{P}{\lambda_p},
\end{align}
which is the SNR when those users work together with ideal cooperation.

\section{Cooperation Bandwidth}
In the above, we have assumed that the RSI can be perfectly shared by all users. In practical systems, the RSI should be first quantized and then sent to the other users via specific D2D cooperation links, which will not only result in an extra quantization noise but also consume extra cooperation bandwidth. Based on the relation between the quantization noise and the cooperation bandwidth, we will find how much bandwidth is required to achieve satisfied performance in this section.\par

To address this issue, we consider a practical RSI sharing, where the receive signals, $y_p$'s, are quantized before sending to the other users. In practical systems, the real and imaginary parts of the received signal should be quantized separately. The overall quantization error can be therefore given by
\begin{align}
\widetilde{y}_p=\mathrm{Re}\{\widetilde{y}_p\}+j\mathrm{Im}\{\widetilde{y}_p\},
\end{align}
If a uniform quantization with $c$ bits are used to quantize $y_p$ at the $p$-th user, then $c/2$ bits are used for the real part and the other bits are for imaginary part. In this case, the variances of the real and the imaginary quantization errors can be given as \cite{BWidrow}
\begin{align}
\mathrm{E}(|\mathrm{Re}\{y_p\}|^2)=\mathrm{E}(|\mathrm{Im}\{y_p\}|^2)=\frac{\tau^2}{3\cdot 2^c},
\end{align}
where $\tau$ denotes the maximum value of either $\mathrm{Re}\{y_p\}$ or $\mathrm{Im}\{y_p\}$, that is, $|\mathrm{Re}\{y_p\}|\leq \tau$ and $|\mathrm{Im}\{y_p\}|\leq \tau$ for $p=1,2,\cdots,P$. As a result, the variance of the overall quantization error is given by
\begin{align}\label{sigma_Q}
\sigma_Q^2=\mathrm{E}(|\widetilde{y}_p|^2)=\frac{2\tau^2}{3\cdot 2^c},
\end{align}
where we have assumed that the quantization errors of the real and the imaginary parts are independent. After that, the quantized signals are sent to the other users. In this situation, the RSI at the $p$-th user can be represented by
\begin{align}
\mathbf{y}+\widetilde{\mathbf{y}}^{(p)},
\end{align}
where $\widetilde{\mathbf{y}}^{(p)}=(\widetilde{y}_1,\cdots,0,\cdots,\widetilde{y}_P)^{\mathrm{T}}$ with the $p$-th entry being zero because the $p$-th user knows its own RSI and thus no quantization needed.\par

In this situation, the recovered symbol at the $p$-th user can be represented by
\begin{align}
\widehat{x}_p&=\mathbf{q}_p^{\mathrm{H}}(\mathbf{y}+\widetilde{\mathbf{y}}^{(p)})\nonumber\\
&=\mathbf{q}_p^{\mathrm{H}}\mathbf{H}_{\mathrm{e}}^{\mathrm{H}}\mathbf{V}\mathbf{x}+\mathbf{q}_p^{\mathrm{H}}\mathbf{z}+\mathbf{q}_p^{\mathrm{H}}\widetilde{\mathbf{y}}^{(p)}.
\end{align}
Compared to (\ref{coop_recv}), the quantization leads to an extra additive noise, and thus the overall noise power in this case is
\begin{align}
N_a=N_0+(1-|q_p[p]|^2)\sigma_Q^2,
\end{align}
where $q_p[p]$ denotes the $p$-th entry of $\mathbf{q}_p$. If we further assume the decoding vector is with constant amplitude, then $|q_p[p]|^2=1/P$. As a result, the overall noise power can be rewritten by
\begin{align}
N_a=N_0+\sigma_Q^2\left(\frac{P-1}{P}\right).
\end{align}\par


In above, we have used $c$ bits for quantization at each user. Alternatively, the number of quantization bits can be also viewed as the amount of information that needs to be shared by the corresponding user. For a practical downlink transmission based on the orthogonal-frequency-division-multiplexing (OFDM) with $K$ subcarriers and symbol duration, $T$, each subcarrier will require $c$ bits for quantization, and thus the required information rate for RSI sharing is
\begin{align}
\frac{cK}{T}=cW~\mathrm{(bit/s)},
\end{align}
where $W=K/T$ is the bandwidth of the downlink transmission. On the other hand, if the bandwidth for cooperation is $W_{\mathrm{c}}$, we should have
\begin{align}\label{rate_req}
cW\leq W_{\mathrm{c}}\log_2(1+\gamma),
\end{align}
with $\gamma$ indicating the SNR of the D2D cooperation link, because the information rate of the D2D cooperation link should be larger than the required rate.\par
Substituting (\ref{rate_req}) into (\ref{sigma_Q}), we can obtain
\begin{align}\label{exp_W}
\sigma_Q^2\propto(1+\gamma)^{-\frac{W_c}{W}}.
\end{align}
Equation (\ref{exp_W}) shows that the power of the quantization error can be reduced exponentially by increasing the cooperation bandwidth. On the other hand, the quality of the D2D cooperation link determines the reduction rate of the quantization error power. As a result, the overall noise power can be reduced with more cooperation bandwidth, and thus the system can be improved accordingly.
\section{Simulation Results}

\begin{figure}
  \centering
  \includegraphics[width=3.8in]{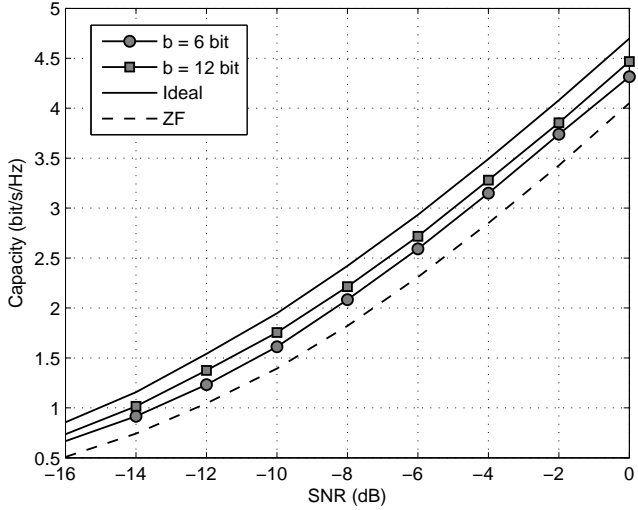}\\
  \caption{Capacities for different approaches versus SNR.}\label{fig1}
\end{figure}
\begin{figure}
  \centering
  \includegraphics[width=3.8in]{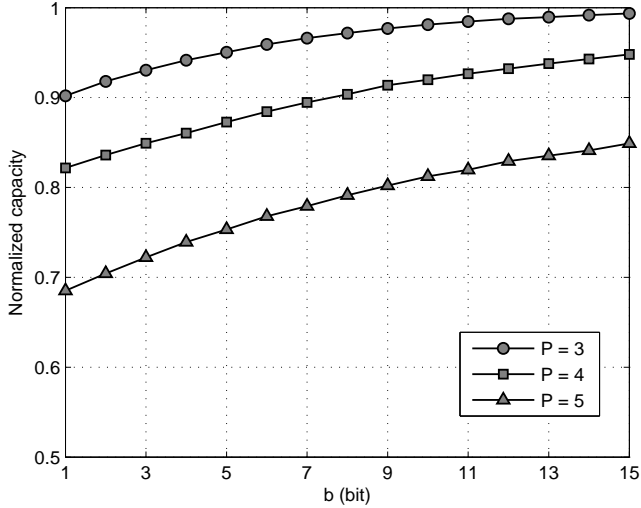}\\
  \caption{Normalized capacities with respect to number of quantization bits.}\label{fig2}
\end{figure}
\begin{figure}
  \centering
  \includegraphics[width=3.8in]{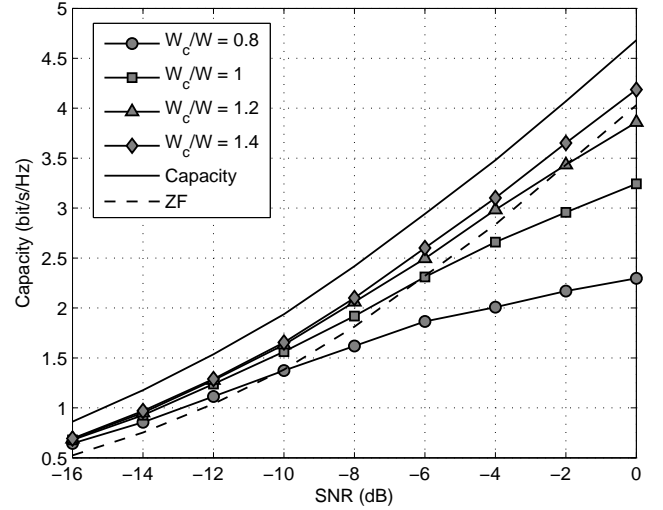}\\
  \caption{Capacities in the presence of the quantization errors.}\label{fig3}
\end{figure}
\begin{figure}
  \centering
  \includegraphics[width=3.8in]{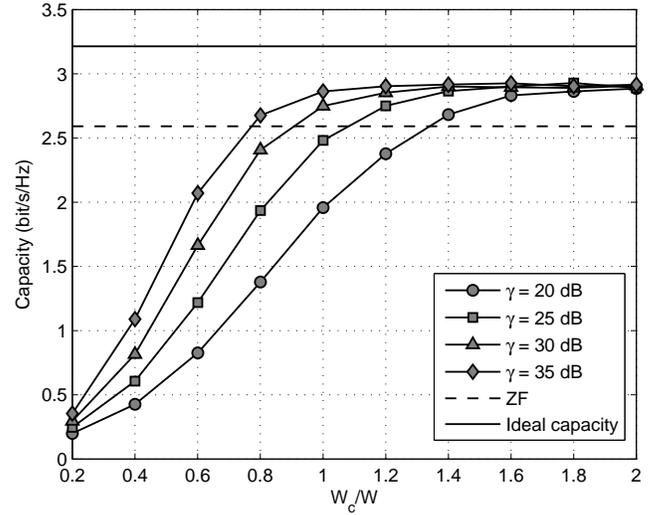}\\
  \caption{Capacities versus cooperation bandwidth with respect to different D2D link quality.}\label{fig4}
\end{figure}

In this section, computer simulation is used to demonstrate the proposed approach. In the simulation, we consider a uniform-linear array (ULA) with $M=64$ antennas and the antenna spacing is half wavelength. In this case, the physical channel can be represented by \cite{AMSayeed}
\begin{align}
\mathbf{h}_p=\sum_{l=0}^{L-1}\alpha_{l,p} \mathbf{s}(\theta_{l}),
\end{align}
where $\mathbf{s}(\theta_l)=(1,e^{j\pi \sin\theta_{l}},\cdots,e^{j\pi \sin\theta_{l}(M-1)})^{\mathrm{T}}$ denotes the steering vector with $\theta_{l}$ indicating the angle-of-arrival of the $l$-th path, and $\alpha_{l,p}$ denotes the complex amplitudes for the $l$-th path corresponding to the $p$-th user with $\mathrm{E}(|\alpha_{l,p}|^2)=1/L$. We consider $L=20$ in the simulation since it is enough to model a practical wireless channel \cite{CXiao}. The dimension of the effective channel is with $D=6$. Without specification, $P=4$ users are considered in the simulation, each of which has a single receive antenna. The users are assumed to have the same spatial covariance matrix and they are also very close to each other such that D2D cooperation can be conducted to share RSI. For comparison, traditional non-cooperative ZF precoding and the case where the users can ideally cooperate with each other are also taken into account. We consider two cases, depending on whether the RSI is ideal.\par
\subsection{Ideal RSI}
Fig.~\ref{fig1} shows the capacities of different approaches versus SNR, where we assume that each user has the ideal RSI. In the figure, we consider $b=6,12$ for decoding codebook design, respectively. As expected, the performance of the proposed approach can be improved compared to the non-cooperative ZF precoding due to the user cooperation. Moreover, the performance can be also improved by using a decoding codebook with a larger size.\par

Fig.~\ref{fig2} presents the performances when $\mathrm{SNR}=-5\text{~dB}$ with respect to the number of quantization bits, where the capacities have been normalized by that of the ideal cooperation case. In this figure, we consider different user numbers with $P=3,4,\text{~and~}5$, respectively. The figure shows that a small number of quantization bits can almost achieve the performance of the ideal cooperation when the user number is small. When the user number is larger, however, much more quantization bits will be required to achieve the ideal one. Actually, such observation coincides with our analysis in Section IV. Equation (\ref{Jindalform}) shows that the bit number is divided by the user number minus one, and therefore more bits will be required when the user number is large.

\subsection{Non-Ideal RSI}
From (\ref{sigma_Q}), the performance depends heavily on $\tau$. By running the simulation many times, we observe that $\tau=30$ is large enough to cover all the received signals.\par

Fig.~\ref{fig3} shows the capacities in the presence of the quantization errors with respect to different cooperation bandwidths. When the cooperation bandwidth is small, the quantization error is dominant and thus the performance can be hardly improved by increasing the SNR. On the other hand, the quantization error can be omitted when the cooperation bandwidth is large and thus the performance can achieve that with ideal RSI case.

Fig.~\ref{fig4} presents the capacities versus the cooperation bandwidth when $\mathrm{SNR}=-5\text{~dB}$ with respect to different D2D link quality. When $\gamma$ is larger, the quality of D2D cooperation link is good and thus a small cooperation bandwidth can already achieve better performance than the non-cooperative ZF. On the other hand, more cooperation bandwidth will be required if the D2D cooperation link is with low quality. Fig.~\ref{fig4} also shows that the required cooperation bandwidth can be even smaller than the downlink transmission bandwidth when the D2D cooperation link is with high quality. In practical systems, the users are supposed to be very close to each and thus a good D2D link quality can be always expected. In this sense, the proposed approach is expected to achieve efficient cooperation with small consumption of extra cooperation bandwidth.

\section{Conclusion}
In this paper, we have investigated the user cooperation in massive MIMO systems with cascaded precoding. If the users can be served by the same outer precoder, they should have the same spatial covariance matrix and thus are supposed to be close to each other. In this situation, D2D enabled user cooperation can be used to improve the system performance. To reduce the amount of the information that needs to be shared, a decoding codebook based approach has been developed such that we only need to share the RSI. The required cooperation bandwidth has also been discussed in the paper. Our simulation results have shown that the proposed approach can achieve cooperation with small consumption of extra cooperation bandwidth.

\bibliographystyle{IEEEtran}
\bibliography{IEEEabrv,lsmimobib}

\begin{thebibliography}{10}
\providecommand{\url}[1]{#1}
\csname url@samestyle\endcsname
\providecommand{\newblock}{\relax}
\providecommand{\bibinfo}[2]{#2}
\providecommand{\BIBentrySTDinterwordspacing}{\spaceskip=0pt\relax}
\providecommand{\BIBentryALTinterwordstretchfactor}{4}
\providecommand{\BIBentryALTinterwordspacing}{\spaceskip=\fontdimen2\font plus
\BIBentryALTinterwordstretchfactor\fontdimen3\font minus
  \fontdimen4\font\relax}
\providecommand{\BIBforeignlanguage}[2]{{%
\expandafter\ifx\csname l@#1\endcsname\relax
\typeout{** WARNING: IEEEtran.bst: No hyphenation pattern has been}%
\typeout{** loaded for the language `#1'. Using the pattern for}%
\typeout{** the default language instead.}%
\else
\language=\csname l@#1\endcsname
\fi
#2}}
\providecommand{\BIBdecl}{\relax}
\BIBdecl

\bibitem{FRusek}
F.~Rusek, D.~Perrsson, B.~K. Lau, E.~G. Larsson, T.~L. Marzetta, O.~Edfors, and
  F.~Tufvesson, ``Scaling up {MIMO}: opportunities and challenges with very
  large arrays,'' \emph{IEEE Signal Process. Mag.}, vol.~30, no.~1, pp. 40--60,
  Jan. 2013.

\bibitem{LLu}
L.~Lu, G.~Y. Li, A.~L. Swindlehurst, A.~Ashikhmin, and R.~Zhang, ``An overview
  of massive {MIMO}: benifits and chanllenges,'' \emph{IEEE J. Sel. Topics
  Signal Process.}, vol.~8, no.~5, pp. 742--758, Oct. 2014.

\bibitem{YLiu_survey}
Y.~Liu, Z.~Tan, H.~Hu, J.~L.~J.~Cimini, and G.~Y. Li, ``Channel estimation for
  {OFDM},'' \emph{IEEE Commun. Survey Tuts.}, vol.~16, no.~4, pp. 1891--1980,
  Fourth Quarter 2014.

\bibitem{DJLove}
D.~J. Love, R.~W. $\text{Heath, Jr.}$, V.~K.~N. Lau, and D.~Gesbert, ``An
  overview of limited feedback in wireless communication systems,'' \emph{IEEE
  J. Sel. Area Commun.}, vol.~26, no.~8, pp. 1341--1365, Oct. 2008.

\bibitem{AAdhikary}
A.~Adhikary, J.~Nam, J.~Y. Ahn, and G.~Caire, ``Joint spatial division and
  multiplexing-the large-scale array regime,'' \emph{IEEE Trans. Inf. Theory},
  vol.~59, no.~10, pp. 6441--6463, Oct. 2013.

\bibitem{JNam}
J.~Nam, J.~Y. Ahn, A.~Adhikary, and G.~Caire, ``Joint spatial division and
  multiplexing: Realizing massive mimo gains with limited channel state
  information,'' in \emph{46-th Information Sciences and Systems (CISS)}, Mar.
  2012, pp. 1--6.

\bibitem{IViering}
I.~Viering, H.~Hofstetter, and W.~Utschick, ``Validity of spatial covariance
  matrices over time and frequency,'' in \emph{Proc. IEEE Globe. Telecommun.
  Conf. (GLOBECOM)}, Taipeh, Taiwan, Nov. 2002, pp. 851--855.

\bibitem{RHClarke}
R.~H. Clarke, ``A statistical theory of mobile-radio reception,'' \emph{Bell
  Syst. Tech. J.}, pp. 957--1000, July-Aug. 1968.

\bibitem{AAsadi}
A.~Asadi, Q.~Wang, and V.~Mancuso, ``A survey on device-to-device comunication
  in celular networks,'' \emph{IEEE Commun. Survey Tuts.}, vol.~16, no.~4, pp.
  1801--1819, Apr. 2014.

\bibitem{DFeng}
D.~Feng, L.~Lu, Y.~Y. Wu, G.~Y. Li, G.~Feng, and S.~Li, ``Device-to-device
  communications underlaying cellular networks,'' \emph{IEEE Trans. Commun.},
  vol.~61, no.~8, pp. 3541--3551, Aug. 2013.

\bibitem{DTse}
D.~Tse and P.~Viswanath, \emph{Fundamentals of Wireless Communication}.\hskip
  1em plus 0.5em minus 0.4em\relax Cambridge University Press, 2005.

\bibitem{ASendonaris}
A.~Sendonaris, E.~Erkip, and B.~Aazhang, ``User cooperation diversity-part ii:
  Implementations aspects and performance analysis,'' \emph{IEEE Trans.
  Commun.}, vol.~51, no.~11, pp. 1939--1948, Nov. 2003.

\bibitem{HJu}
\BIBentryALTinterwordspacing
H.~Ju and R.~Zhang, ``User cooperation in wireless powered communication
  networks.'' [Online]. Available: \url{http://arxiv.org/abs/1403.7123.}
\BIBentrySTDinterwordspacing

\bibitem{TEHunter}
T.~E. Hunter and A.~Nosratinia, ``Distributed protocols for user cooperation in
  multi-user wireleess networks,'' in \emph{Proc. IEEE Globe. Telecommun. Conf.
  (GLOBECOM)}, 2004, pp. 3788--3729.

\bibitem{NRavindran}
N.~Ravindran and N.~Jindal, ``Limited feedback-based block diagonalization for
  the {MIMO} broadcast channel,'' \emph{IEEE J. Sel. Area Commun.}, vol.~26,
  no.~8, pp. 1473--1482, Oct. 2008.

\bibitem{DJLove_Unitary}
D.~J. Love and R.~W. $\text{Heath, Jr.}$, ``Limited feedback unitary precoding
  for spatial multiplexing systems,'' \emph{IEEE Trans. Inf. Theory}, vol.~51,
  no.~8, pp. 2967--2976, Aug. 2005.

\bibitem{SBoyd}
S.~Boyd, \emph{Convex Optimization}.\hskip 1em plus 0.5em minus 0.4em\relax
  Cmbridge University Press, 2004.

\bibitem{NJindal}
N.~Jindal, ``{MIMO} broadcast channels with finite-rate feedback,'' \emph{IEEE
  Trans. Inf. Theory}, vol.~52, no.~11, pp. 5045 -- 5060, Nov. 2006.

\bibitem{BWidrow}
B.~Widrow, I.~Koll$\acute{\text{a}}$r, and M.~Liu, ``Statistical theory of
  quantization,'' \emph{IEEE Trans. Instrum. Meas.}, vol.~45, no.~2, pp.
  353--361, Apr. 1996.

\bibitem{AMSayeed}
A.~M. Sayeed, ``Deconstructing multiantenna fading channels,'' \emph{IEEE
  Trans. Signal Process.}, vol.~50, no.~10, pp. 2563--2579, Oct. 2002.

\bibitem{CXiao}
C.~Xiao, Y.~R. Zheng, and N.~C. Beaulieu, ``Novel sum-ofsinusoids simulation
  models for rayleigh and rician fading channels,'' \emph{IEEE Trans. Wirel.
  Commun.}, vol.~5, no.~12, pp. 3667--3679, Dec. 2006.

\end{thebibliography}

\end{document}